\documentclass[
aps, 
prl,
preprintnumbers,
twocolumn,
superscriptaddress,
nofootinbib,
floatfix,
10pt
]{revtex4-2}

\usepackage{graphicx}
\usepackage{dcolumn}
\usepackage{bm}
\usepackage{float}
\usepackage{braket}
\usepackage{amsmath}
\usepackage{amssymb}
\usepackage{mathtools}
\usepackage{comment}
\usepackage{xcolor}
\usepackage[bottom]{footmisc}
\usepackage{hyperref}

\def\orcid#1{\kern .08em\href{https://orcid.org/#1}{\includegraphics[keepaspectratio,width=0.7em]{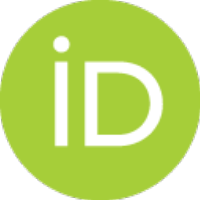}}}

\newcommand{\defeq}{\vcentcolon=}

\newcommand{\IntName}{$\Delta$NNLO${}_{\mathrm{GO}}(394)$}
\newcommand{\MeV}{\,\mathrm{MeV}}

\begin{document}

\title{\textbf{Time-ordered Diagrammatic Monte Carlo for atomic nuclei} 
}%

\author{Stefano~Brolli\orcid{0009-0005-8290-8462}}
 \email{Stefano.Brolli@unimi.it}
\affiliation{Dipartimento di Fisica “Aldo Pontremoli,” Universit\`a degli Studi di Milano, via Celoria 16, I-20133 Milano, Italy}
\affiliation{INFN, Sezione di Milano, via Celoria 16, I-20133 Milano, Italy}

\author{Carlo~Barbieri\orcid{0000-0001-8658-6927}}
 \email{Carlo.Barbieri@unimi.it}
\affiliation{Dipartimento di Fisica “Aldo Pontremoli,” Universit\`a degli Studi di Milano, via Celoria 16, I-20133 Milano, Italy}
\affiliation{INFN, Sezione di Milano, via Celoria 16, I-20133 Milano, Italy}

\date{\today}

\begin{abstract}
\noindent Diagrammatic Monte Carlo provides a systematically improvable framework for stochastically resumming many-body expansions to high orders through direct sampling of diagram topologies. We advance our earlier work by introducing a novel time-ordered Diagrammatic Monte Carlo algorithm for the single-particle Green’s function. The algorithm is tailored to finite nuclei, formulated in discrete model spaces and applicable to arbitrary two-body interactions. The new time-ordered diagrammatic Monte Carlo algorithm is based on the on-the-fly evaluation of time-ordered Goldstone diagrams, avoiding explicit diagram enumeration and expensive frequency integration. We show the algorithm by computing $^{16}$O up to fifth order in a reduced model space using optimized reference state orbitals and including effective three-body forces. Benchmarking against established truncation schemes in \emph{ab initio} nuclear theory demonstrates its potential to overcome the limitations of current many-body approaches.
\end{abstract}

\maketitle

Methods for solving the many-body Schrödinger equation are central to quantum many-body physics, spanning condensed matter, quantum chemistry, and nuclear physics. Their development has therefore been a sustained cross-disciplinary effort, with advances in one field propagating to others. Among the available approaches, those based on diagrammatic expansions have proven particularly powerful, offering a systematically improvable framework together with an intuitive graphical representation of many-body correlations. However, their evaluation at high orders remains a major computational challenge. \emph{Ab initio} nuclear theory provides a stringent testbed for such developments. Atomic nuclei are strongly correlated fermionic systems in which mean-field descriptions are often insufficient, placing high demands on many-body methods. Recent advances enable calculations of heavy~\cite{Bonaiti2025, Hu2022} and deformed~\cite{Sun2025, Hu2024} nuclei within low-order truncation schemes, where ground-state energies and densities are typically well converged. However, next-generation chiral interactions~\cite{Hu2025, Arthuis2024, Jiang2020, Ekstrom2015} require a more accurate and systematically improvable treatment of correlations. Moreover, emergent structure and reaction observables already demand the inclusion of high-order virtual excitations. Examples include quartet formation near the neutron drip line~\cite{Ma2026}, $\alpha$ clustering~\cite{Shen2023, Epelbaum2012}, and cross sections in medium- and heavy-mass nuclei, which are sensitive to many-particle–many-hole configurations beyond currently accessible orders~\cite{Idini2019, Rotureau2018}. The need for controlled high-order many-body methods is further amplified by the precision data produced at radioactive-ion-beam facilities~\cite{FRIB, SPES, FAIR, ISOLDE} and by applications to searches for physics beyond the Standard Model, including CKM unitarity tests~\cite{Hardy2020}, nuclear electric dipole moments~\cite{Engel2013}, and neutrinoless double-beta decay~\cite{Belley2024, Belley2021}. A promising route toward this goal is the combination of many-body perturbation theory with stochastic sampling. Recent work has coupled automated diagrammatic generators~\cite{ArthuisADG2_2021, ArthuisADG1_2019} to Monte Carlo techniques, enabling diagram-by-diagram evaluations of high-order contributions in infinite nuclear matter up to fifth order~\cite{Drischler2026, Drischler2019}. Here we pursue a complementary strategy: instead of explicitly enumerating all diagrams, we sample the diagrammatic space directly. Diagrammatic Monte Carlo (DiagMC), originally developed in condensed matter physics~\cite{Luo2025, VanHoucke2019, VanHoucke2012, Prokofev1998}, generates diagrams on the fly and samples topologies, time orderings, and internal quantum numbers within a single Markov chain. Its extension to finite nuclei at zero temperature requires dedicated algorithms, but offers a route to high-order perturbative expansions without explicit topology enumeration. A first proof-of-principle step in this direction was demonstrated in the Richardson pairing model~\cite{Brolli2025}.
\begin{figure}[t]
    \centering
    \includegraphics[width=\linewidth]{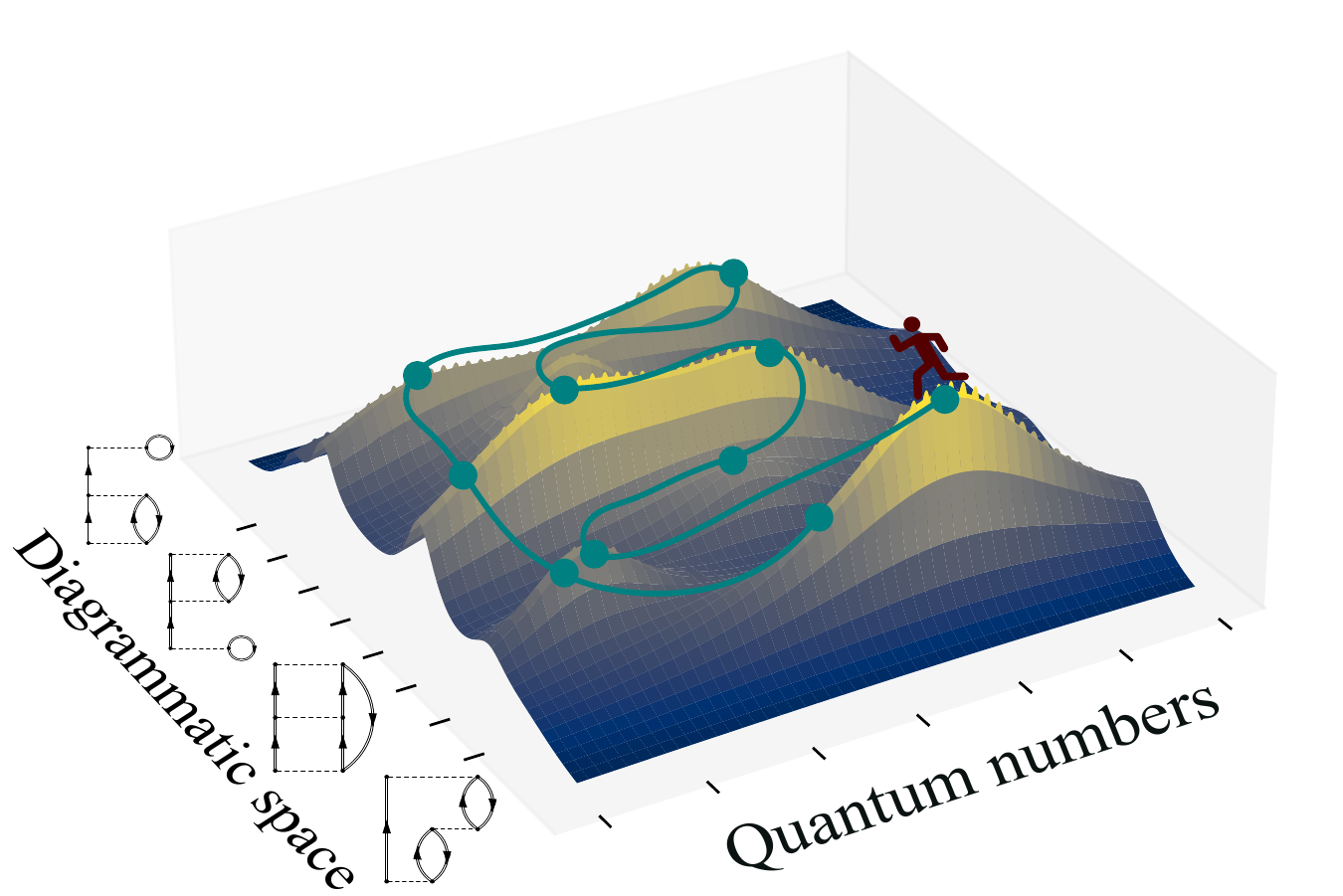}
    \caption{Visual representation of a walker moving in diagrammatic space.}
    \label{fig:DiagMC}
\end{figure}

In this work, we build on this approach by applying DiagMC within the self-consistent Green’s function (SCGF) framework to the $^{16}\mathrm{O}$ nucleus. The Green's function $G(\omega)$ is obtained by solving the Dyson equation $G(\omega) = G^{(0)}(\omega) + G^{(0)}(\omega) \Sigma^\star(\omega)G(\omega)$, where $G^{(0)}$ is the unperturbed propagator and $\Sigma^\star(\omega)$ is the irreducible self-energy. Within SCGF theory, the self-energy corresponds to the microscopic optical potential \cite{Capuzzi1996, Mahaux1995}, naturally connecting structure and reaction observables.

In this Letter, we present a novel algorithm referred to as time-ordered Diagrammatic Monte Carlo (TO-DiagMC) that significantly advances earlier DiagMC implementations. In particular, the method performs a stochastic resummation of Goldstone (time-ordered) diagrams, avoiding the computationally expensive stochastic integration over frequencies. This enables a more accurate and systematic evaluation of the self-energy, substantially reducing the computational cost and allowing calculations up to the fifth order.

\paragraph{\itshape Time-ordered Diagrammatic Monte Carlo} --- 
\begin{figure}
    \centering
    \includegraphics[width=\linewidth]{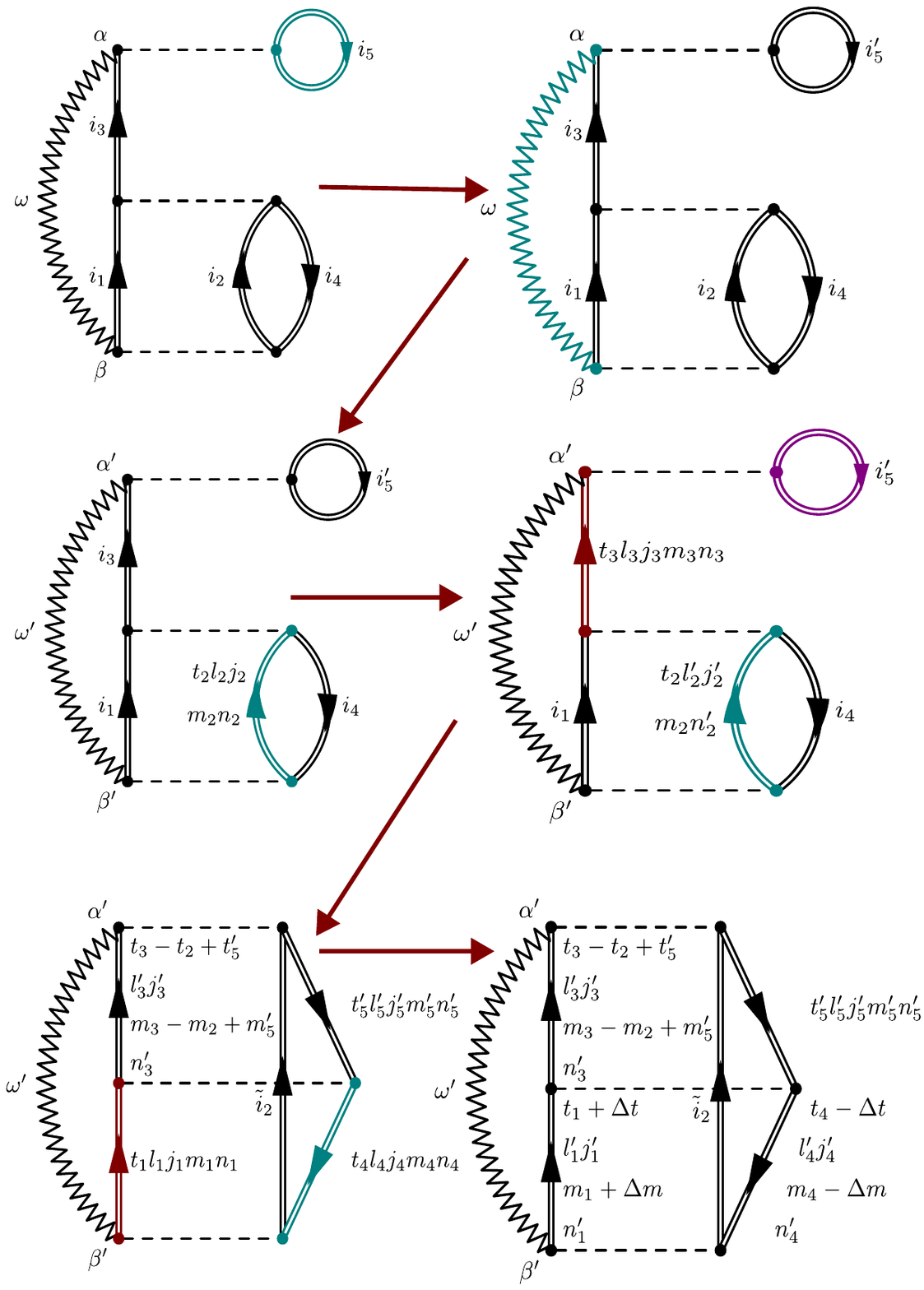}
    \caption{Possible sequence of updates through which a third-order diagram evolves from a non-skeleton topology into a skeleton ring \cite{note:ring, SupplMat}. The path showcases all updates employed in the simulations. The zigzag line connecting the outgoing and incoming vertices of each diagram, referred to as the ``measuring line,'' is required to assign the external frequency $\omega$ within the Goldstone rules. In the figure, we highlight the propagators that are reconnected by each update or whose quantum numbers are modified.}
    \label{fig::updates}
\end{figure}
To improve numerical efficiency and accelerate convergence, we partition the self-energy into frequency bins and, within each bin, project it onto a basis of Legendre polynomials. The associated projection integrals are evaluated using Gauss–Legendre quadrature. The quadrature nodes and corresponding weights are sampled with the topologies and quantum numbers \cite{SupplMat}. DiagMC then recasts the diagrammatic series of the self-energy as a weighted average of the signs (or phases) of the diagrams \cite{VanHoucke2019}. The choice of the weight function is critical in controlling the sign problem of the diagrammatic series \cite{Luo2025}. Our choice of weight function was dictated by the following observations. Given a Hamiltonian with a two-body potential $H = H^{(0)} + V$, a Feynman diagram of order $n$ can be schematically expressed as

\begin{equation} \label{eq1}
\mathcal{D}_{\alpha \beta}^{(n)} (\omega) = \sum_{\{a\}} \left[ \prod_{v = 1}^n V_v  \right] \int \prod_{l = 1}^n \frac{d\omega_l}{2 \pi} \left[ \prod_{r = 1}^{2n - 1} G^{\mathrm{RS}}_{a_r}(\omega_r) \right]
\end{equation}

\noindent where $V_v \defeq V_{a_{v_1} a_{v_2} a_{v_3} a_{v_4}}$ and $\alpha, \beta$ are the only two uncontracted indices in the interactions $V_v$. The frequencies $\omega_r$ are constrained by energy conservation at each vertex. 
The reference state (RS) propagator $G^{\mathrm{RS}}$ can be made diagonal by dressing the interaction as
$V_{\alpha\beta\gamma\delta} \rightarrow \sum_{\alpha \beta \gamma \delta} V_{\alpha\beta\gamma\delta}\,
(\mathcal{Z}^{i_1}_{\alpha}\mathcal{Z}^{i_2}_{\beta})^*
\mathcal{Z}^{i_3}_{\gamma}\mathcal{Z}^{i_4}_{\delta}$, where $\mathcal{Z}$ denotes the spectral amplitudes appearing in the numerators of the RS propagator~\cite{Barbieri::LectNotesPhys936}. This dressing reduces the space that needs to be explored by the DiagMC simulation and makes the RS propagator a function with a single pole, which is easier to treat analytically. 
In our simulation, we employ a Hartree-Fock (HF) reference state propagator. This contraction amounts to working in the HF basis; however, it is more general as it can be applied to a fully dressed propagator. We enforce full self-consistency on the static part of the self-energy with the so-called sc0 approximation \cite{Soma2014::Gorkov, Barbieri2022::GorkovADC(3)}.
The frequency integrals in Eq.~\eqref{eq1} can be evaluated using Goldstone’s rules~\cite{Schirmer2018LNC}, which effectively implement the residue theorem for functions with poles in the complex plane. In this framework, each allowed time ordering of the interaction vertices in a Feynman diagram yields a distinct contribution, leading to a sum of the form $\sum_T R_T^v(\omega)$, where $T$ runs over all time orderings compatible with the particle–hole structure of the interaction, and $v$ collectively denotes the quantum numbers associated with the interaction vertices. Equation~\eqref{eq1} can then be recast in the DiagMC-friendly form

\begin{equation} \label{eq2}
\begin{split}
\mathcal{D}_{\alpha \beta}^{(n)} (\omega) & = \sum_{\{a\}} \sum_{T} \left\lvert \prod_{v = 1}^n V_v \right\rvert \mathrm{sign} \left( \prod_{v = 1}^n V_v \right) p_T \frac{R_T^v(\omega)}{p_T} \\
& = \lim_{N \to \infty} \frac{Z_{\alpha\beta}}{N} \sum_j^N \mathrm{sign}\left( \prod_{v = 1}^n V_{v_j} \right) \frac{R_{T_j}^{v_j}(\omega_j)}{p_{T_j}}.
\end{split}
\end{equation}

\noindent The last equality represents a stochastic average of diagrams generated with the probability distribution function
\begin{equation} \label{eq_weight}
w_T^{v} = \frac{1}{Z_{\alpha\beta}} \left|\prod_{v=1}^{n} V_v\right|\, p_T ,
\end{equation}
where $p_T$ is the probability of choosing a specific Golgstone time ordering, and $Z_{\alpha\beta}$ is a normalization constant. 
The diagrams are sampled using a Monte Carlo Markov chain in the space of diagrams. The Markovian walkers evolve in a high-dimensional space of topologies, time orderings, and quantum numbers, for which Fig.~\ref{fig:DiagMC} provides a pictorial representation. Each possible ``move" is described by an ``update", which acts as the proposal function of the Metropolis-Hastings algorithm \cite{Hastings1970, Metropolis1953}. The updates can change the topology of the diagram, its quantum numbers, or both at the same time. The requirement to reproduce the target probability distribution~\eqref{eq_weight} is met through ergodicity and tuned acceptance ratios.
We performed a different simulation for each order using the five updates shown in Fig. \ref{fig::updates}. We also explicitly checked ergodicity on the diagram topologies up to order four. More details on the algorithm and the Markov chain updates are available in Supplemental Material \cite{SupplMat}. 

\begin{figure}
    \centering
    \includegraphics[width=\linewidth]{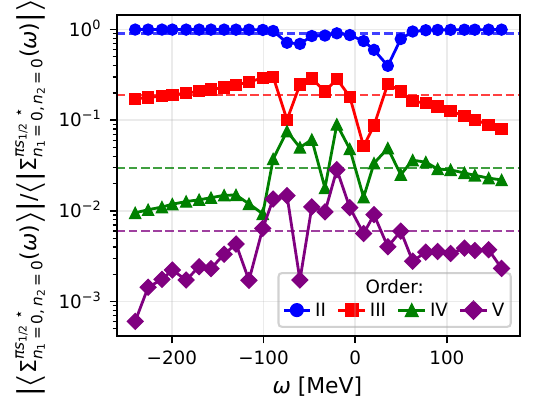}
    \caption{Average sign of the self-energy contributions for several values of $\omega$ of different perturbative orders for the component $n_1=n_2=0$ of the protonic $s_{1/2}$ wave in the HF basis and $\eta = 10 \MeV$. The dashed lines represent the average sign contribitions for diagrams at each order.}
    \label{fig:sign_problem}
\end{figure}

The TO-DiagMC algorithm eliminates the sign problem associated with frequency integrations by removing the frequency integration sector altogether. Figure~\ref{fig:sign_problem} illustrates the severity of the sign problem at selected nodes of the Gauss–Legendre $\omega$ integration grid. TO-DiagMC nearly eliminates the sign problem at second order and reduces it to manageable levels up to fifth order. Extending TO-DiagMC to larger model spaces is the central final objective. Achieving this requires optimizing the sampling weight to mitigate the explicit sign dependence, $\mathrm{sign}\left(\prod_{v=1}^{n} V_{v_j}\right)$ in Eq.~\eqref{eq2}. A promising strategy is the on-the-fly summation over groups of quantum numbers, enabling the explicit cancellation of sign-alternating contributions within each sampled diagram.

\paragraph{\itshape Model} --- 
We performed TO-DiagMC simulations for the ${}^{16}\mathrm{O}$ nucleus in a reduced model space with $e_{\max}=2$, employing an optimized RS (OpRS) basis and the soft chiral interaction \IntName \cite{Jiang2020}. The OpRS basis is a natural extension of natural orbitals within the SCGF framework and provides the independent-particle model that best reproduces the lower moments of the spectral strength distribution \cite{Marino2026, Barbieri2022::GorkovADC(3), Barbieri2009}. This has the advantage of accurately reproducing the ground state energy of the system within a mean-field reference state and incorporating, in the most consistent way, the relevant correlations of the system into the reference orbitals. We note that this is the first calculation performed using a reduction induced by OpRS orbitals. The combination of a soft interaction and OpRS orbitals enables the extraction of meaningful information in light nuclei even within truncated model spaces; however, the present application should be regarded as a proof of applicability rather than a calculation aimed at achieving full convergence with respect to the truncation. In this setup, the diagrammatic expansion was sampled up to the fifth order. 
\begin{figure}[t]
    \centering
    \includegraphics[width=\linewidth]{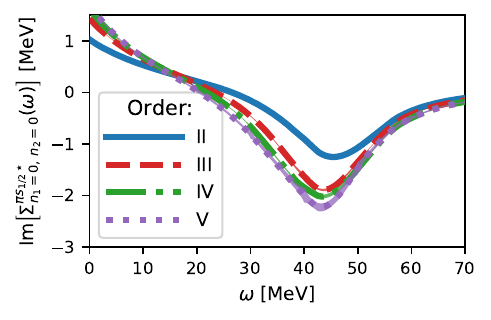}
    \caption{Imaginary part of the $n_1 = n_2 = 0$ component of the self-energy in the proton $s_{1/2}$ partial wave, computed in the HF basis with $\eta = 10 \MeV$. The real part can be reconstructed via a dispersion relation. Monte Carlo uncertainties are shown as a band and are only visible at fifth order.}
    \label{FigSE}
\end{figure}
The simulations require the introduction of a finite regulator $\eta$, since the denominators of the Green's function contain poles with an infinitesimal imaginary part. In the limit $\eta \to 0^+$, the self-energy develops oscillations that become increasingly difficult to control as the perturbative order grows. This behavior originates from the appearance, in the perturbative expansion, of terms scaling as $\sim (V/\eta)^{n-1}$ in the vicinity of the poles, where $n$ denotes the expansion order. A practical way to enforce convergence is to choose $\eta \gtrsim \max(V)$. We have verified that using $\eta = 10\,\mathrm{MeV}$ is sufficient to enforce convergence over the entire energy range. While this choice suppresses fine spectral structures, in this first application of TO-DiagMC, we adopt a large regulator and focus on obtaining a stable self-energy. Importantly, using a large regulator does not affect the normalization of the self-energy; physically, it amounts to grouping many particle-hole excitations within a broadened energy. The controlled extrapolation to the physical limit $\eta \to 0^+$ in the self-energy calculation will be addressed in future work. In this respect, sampling in the Goldstone-diagram space is advantageous, as the ADC coupling and intermediate-state configuration matrices (often called $M$, $N$, $C$, and $D$ \cite{Barbieri::LectNotesPhys936}) appear more explicitly in this representation, enabling easier reorganization of the expansion that mitigates the strong oscillations of bare perturbation theory \cite{Barbieri::LectNotesPhys936, Raimondi2018}.
We note, however, that this is related to the perturbative expansion of the Green's function and not to TO-DiagMC. TO-DiagMC is not intrinsically tied to the presence of a regulator and can, in its present form, be directly applied to perturbative expansions that are well defined without introducing an explicit $\eta$, such as many-body perturbation theory.

\paragraph {\itshape Results} ---  
\begin{figure}[t]
    \centering
    \includegraphics[width=\linewidth]{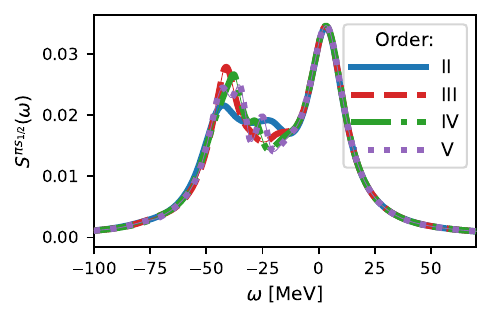}
    \includegraphics[width=\linewidth]{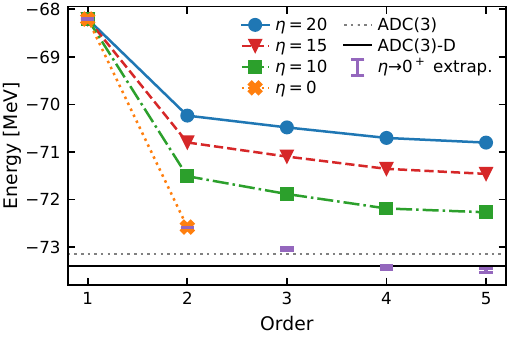}
    \caption{Top: Proton $s_{1/2}$ spectral function computed at successive perturbative orders within TO-DiagMC, using a regulator $\eta = 10$ MeV.
    Bottom: Ground-state energy as a function of perturbative order in the self-energy for different values of the regulator $\eta$. Results at $\eta = 20$, $15$, and $10 \MeV$ are obtained from TO-DiagMC simulations, while the second-order point at $\eta = 0$ is obtained with the SCGF framework of Refs.~\cite{Barbieri::LectNotesPhys936, Cipollone2013}. Extrapolations to $\eta \to 0^+$ are performed from the TO-DiagMC data as described in the main text.}
    \label{FigE_and_SF}
\end{figure}
In Fig.~\ref{FigSE}, we present the $n_1 = n_2 = 0$ component (with $n$ the principal quantum number) of the dynamical self-energy in the proton $s_{1/2}$ partial wave and the HF basis, which constitutes the dominant contribution in this channel. The self-energy is computed using TO-DiagMC and $\eta = 10 \MeV$ and shown over the energy range relevant for optical potentials, $\omega \in [0, 70] \MeV$. The Monte Carlo uncertainty is estimated as the standard deviation of the mean over ten independent runs. Each run employs 128 walkers evolving in parallel and performing $\sim 10^{11}$ updates in diagrammatic space. As shown, the resulting statistical error is barely visible just at fifth order and remains significantly smaller than the differences between successive perturbative orders. Moreover, it is subdominant compared to the uncertainty associated with the finite regulator $\eta$. In the following, Monte Carlo statistical errors are therefore neglected.
The figure shows that clear differences emerge between second- and fifth-order results, underscoring the significant impact of higher-order correlations on the dynamical structure of the optical potential. These effects arise from the progressive inclusion of many-particle–many-hole virtual excitations, which are essential to capture absorption mechanisms associated with coupling to inelastic channels. Consistently, higher-order contributions lead to a more negative self-energy, reflecting enhanced absorption. 

Figure~\ref{FigE_and_SF} shows the impact of higher-order contributions on the spectral function and the ground-state energy. The former, shown in the upper panel, encodes the full one-particle spectroscopic information~\cite{Dickhoff:manybody}. It is obtained by solving the Dyson equation on an energy mesh $\omega \in [-240,160]~\MeV$, spanning the region where the dynamical self-energy has significant support. Although the finite regulator $\eta$ smooths fine spectroscopic details, the spectral function displays clear modifications induced by higher-order contributions. The ground-state energy can be obtained from the Green's function through the Galitskii-Migdal-Koltun sum rule~\cite{Migdal1958, Koltun1974, Dickhoff:manybody, SupplMat}, shown in the lower panel of Fig.~\ref{FigE_and_SF}. At second order, the analytic structure of the self-energy is exactly preserved, allowing $\eta$ to be reduced arbitrarily in TO-DiagMC and the $\eta=0$ limit to be taken exactly in standard SCGF approaches. This provides a useful benchmark and a powerful extrapolation technique. First, we verified that decreasing $\eta$ to $0.5\MeV$ in TO-DiagMC systematically brings the result toward the exact limit $\eta = 0^+$ of the standard SCGF method of Refs~\cite{Barbieri::LectNotesPhys936, Cipollone2013}. To extrapolate the TO-DiagMC results to $\eta\to0^+$, we used two different techniques and assigned their difference as the extrapolation uncertainty. First, we performed an offset-corrected linear extrapolation. A linear fit to the second-order results at $\eta=20, 15$, and $10~\MeV$ determines the deviation from the exact $\eta=0$ benchmark; the resulting offset is then used to correct the higher-order linear extrapolations. Second, we used a calibrated power-law form, $E(\eta)=E(0)+A\eta^\alpha$.
At second order, fixing $E(0)$ to the exact SCGF value determines the exponent $\alpha$ from the finite-$\eta$ data. The same exponent is then used at higher orders, where $E(0)$ and $A$ are fitted. Fig.~\ref{FigE_and_SF} shows that increasing the perturbative order systematically brings the TO-DiagMC results closer to the ADC(3)-D calculation, the current state of the art in SCGF theory~\cite{Marino2026, Barbieri::LectNotesPhys936}. While ADC(3) already improves upon strict third-order perturbation theory through additional resummations, the visible fourth- and fifth-order TO-DiagMC contributions indicate that higher-order correlations, better captured by ADC(3)-D, remain important.
These results establish TO-DiagMC as a viable route for systematically accessing high-order contributions to the nuclear self-energy in \emph{ab initio} calculations.

{\itshape Conclusion} --- In this Letter, we demonstrated that TO-DiagMC can resum self-energy diagrams up to fifth order, which has not been achieved previously in nuclear-structure applications. This establishes TO-DiagMC as a promising new tool for \emph{ab initio} nuclear theory. Several developments could further extend the reachable model spaces $e_{\mathrm{max}}$, including the use of GPUs to accelerate the tensor contractions entering the diagrammatic weights, which may also help mitigate the sign problem. Improvements in the extrapolation to $\eta \to 0^+$ will also be important to better resolve the spectral structure. This could be done by combining nonperturbative resummations, which would reduce the number of diagrams and aid sampling. These developments will be explored in future work. Meanwhile, the generality of TO-DiagMC enables immediate applications within frameworks such as many-body perturbation theory, where diagrams are energy independent and no regulator is required.

\paragraph {\itshape Acknowledgments} --- 
The authors thank E. Vigezzi and Dean Lee for reading the manuscript and for the helpful comments and suggestions.
This work used the DiRAC Data Intensive service (DIaL3) at the University of Leicester, managed by the University of Leicester Research Computing Service on behalf of the STFC DiRAC HPC Facility (www.dirac.ac.uk). The DiRAC service at Leicester was funded by BEIS, UKRI and STFC capital funding and STFC operations grants. DiRAC is part of the UKRI Digital Research Infrastructure.

\paragraph {\itshape Data availability} --- The data that support the findings of this article are not publicly available. The data are available from the authors upon reasonable request.

\bibliographystyle{apsrev4-2}
\bibliography{bibliography}

\end{document}


\begin{center}
\section*{Supplemental Material}
\end{center}

\renewcommand{\thefigure}{S\arabic{figure}}
\renewcommand{\thetable}{S\arabic{table}}
\renewcommand{\theequation}{S\arabic{equation}}
\renewcommand{\thepage}{\arabic{page}}
\setcounter{figure}{0}
\setcounter{table}{0}
\setcounter{equation}{0}
\setcounter{page}{1} 

\subsection*{The algorithm}

The time-ordered Diagrammatic Monte Carlo (TO-DiagMC) algorithm computes the self-energy of the single-particle Green's function by sampling time-ordered (Goldstone) diagrams. The self-energy exhibits the asymptotic behavior $\sim A / \omega + i B / \omega^2$, where $A$ and $B$ are real numbers \cite{Dickhoff:manybody}. This ensures the existence of a finite energy window — which can be estimated from the reference propagator — extending over a few hundred MeV, within which the self-energy is appreciably different from zero. To optimize the sampling, we partition this window into bins of width of order $\eta$. Given the relatively large value of $\eta$ used in the present work, this results in only a small number of bins for each matrix element of the self-energy.
Moreover, the self-energy is diagonal in $tljm$ space and independent of $m$. We therefore set $m = -j$ without loss of generality. As a consequence, the self-energy can be computed independently for each partial wave, and the only external quantum numbers that need to be sampled are the incoming and outgoing principal quantum numbers $n_1$ and $n_2$.
To further improve the sampling efficiency, within each bin we expand the self-energy on the basis of the first $N$ normalized Legendre polynomials, up to degree $N-1$. Denoting this basis by $\{ B^{v}(\omega) \}_{v=0}^{N-1}$, we define the projected components

\begin{equation} \label{eqProj}
    \Sigma^{tlj (v)}_{n_1 n_2} = \int d \omega \, B^{v}(\omega) {\Sigma^{\star}}^{tlj}_{n_1n_2} (\omega).
\end{equation}

This expansion assumes that, within each bin, the self-energy is well approximated by a polynomial of degree $N-1$. Consequently, the integrand in Eq.~\eqref{eqProj} is well approximated by a polynomial of degree $2N-2$. Such integrals can therefore be evaluated exactly using Gauss--Legendre quadrature with $N$ nodes, yielding

\begin{equation} \label{eqGaussLeg}
    \Sigma^{tlj (v)}_{n_1 n_2} = \sum_{k = 1}^{N} W_k B^{v}(\omega_k) {\Sigma^{\star}}^{tlj}_{n_1n_2} (\omega_k),
\end{equation}

\noindent where $\omega_k$ and $W_k$ are the Gauss-Legendre nodes and weights, respectively.

Expanding the self-energy in terms of diagrams, we obtain

\begin{equation} \label{eqGaussLeg}
    \Sigma^{tlj (v)}_{n_1 n_2} = \sum_{k = 1}^{N} W_k B^{v}(\omega_k) \sum_{\mathcal{T}} \mathcal{D}_{n_1n_2}^{tlj} (\omega_k, \mathcal{T}) 1_{\mathcal{T} \in S_{\Sigma}},
\end{equation}

\noindent where $S_{\Sigma}$ denotes the set of diagrams contributing to the self-energy, $1_{\mathcal{T} \in S_{\Sigma}}$ is its characteristic function, and $\mathcal{D}^{tlj}_{n_1 n_2}(\omega_k,\mathcal{T})$ represents the contribution of a diagram with topology $\mathcal{T}$ evaluated at frequency $\omega_k$.

The definition of $S_{\Sigma}$ depends on the chosen approximation scheme. In self-consistent calculations (e.g., sc0), it consists of connected, skeleton, one-particle irreducible diagrams. In non-self-consistent approaches, it reduces to the set of connected, one-particle irreducible diagrams.

Each diagram is expressed as a product of interaction matrix elements $V_{\alpha \beta \gamma \delta}$ and propagators $G^{RS}_{\alpha \beta}(\omega)$. The propagators have the spectral representation

\begin{equation} \label{eq_Lehmann_form}
    G_{\alpha \beta}^{RS} (\omega) = \sum_{i} \frac{\mathcal{Z}_\alpha^i \mathcal{Z}^i_{\beta}}{\omega - \epsilon_{i} + i \,\mathrm{sign}(i - F) \, \eta} 
\end{equation}

\noindent where $F$ denotes the Fermi level \cite{Dickhoff:manybody}.

It is convenient to contract the residues with the interaction matrix elements and define  $\tilde{V}_{ijkl} = \sum_{\alpha \beta \gamma \delta} V_{\alpha \beta \gamma \delta} \mathcal{Z}^i_\alpha \mathcal{Z}^j_\beta \mathcal{Z}^k_\gamma \mathcal{Z}^l_\delta$. In this representation the propagators become diagonal,

\begin{equation}
    \tilde{G}_{ij} (\omega) = \frac{\delta_{ij}}{\omega - \epsilon_i +  i \,\mathrm{sign}(i - F) \, \eta}.
\end{equation}

Each index $i,j,k,\dots$ is uniquely associated with either a particle or a hole state. Once these quantum numbers are fixed at each vertex, the integrals over internal frequencies generate only those Goldstone time orderings that are compatible with the overall particle-hole structure; all others vanish identically.

Denoting the non-vanishing time-ordered contributions by $R_T^{\mathcal{T}{i}}(\omega)$, where ${i}$ collectively labels all internal quantum numbers, we obtain

\begin{equation} \label{eqSupp1}
    \Sigma^{tlj (v)}_{n_1 n_2} = \sum_{k = 1}^{N} \sum_{\mathcal{T}} \sum_{\{i\}} \sum_{T} W_k B^{v}(\omega_k) \left[ \prod V_{\{i\}} \right] R_T^{\mathcal{T} \{i\}}(\omega_k) 1_{\mathcal{T} \in S_{\Sigma}},
\end{equation}

\noindent where $\prod V_{{i}}$ denotes the product of all interaction matrix elements in the diagram with indices ${i}$.

Equation~\eqref{eqSupp1} can be cast into a form suitable for DiagMC sampling,

\begin{equation}  \label{eqDiagMC}
    \begin{split}
    \Sigma^{tlj (v)}_{n_1 n_2} & = \sum_{k = 1}^{N} \sum_{\mathcal{T}} \sum_{\{i\}} \sum_{T} W_k B^{v}(\omega_k) \left\lvert \prod V_{\{i\}} \right\rvert  \mathrm{sign} \left( \prod V_{\{i\}} \right) R_T^{\mathcal{T} \{i\}}(\omega_k) 1_{\mathcal{T} \in S_{\Sigma}} \\
    & = Z_{n_1n_2}^{(tlj)} \lim_{N \to \infty} \frac{1}{N} \sum_{r=1}^N B^v \left( \omega_r \right) \mathrm{sign} \left( \prod V_{\{i\}_r} \right) \frac{R_{T_r}^{\mathcal{T}_r \{i\}_r}}{p_{T_r}}(\omega_r) 1_{\mathcal{T}_r \in S_{\Sigma}},
    \end{split}
\end{equation}

\noindent where the second line represents an average over time-ordered diagrams sampled with the probability distribution 

\begin{equation} \label{eqWeight}
    w_{\{i\}k} = {Z_{n_1n_2}^{(tlj)}}^{-1} W_k \left\lvert \prod V_{\{i\}} \right \rvert p_T \mathcal{R}(\mathcal{T}).
\end{equation}

\noindent $\mathcal{R}(\mathcal{T})$ is a reweighting factor that can be used to enhance the sampling of skeleton (or normalization) diagrams with respect to non-skeleton and non-normalization topologies. We have selected $\mathcal{R}(\mathcal{T}) = 1$ if $\mathcal{T} \in S_{\Sigma}$, and $\mathcal{R}(\mathcal{T}) < 1$ for other topologies. $p_T$ denotes the probability of selecting a given time ordering. The selection of the time ordering is implemented through the inequalities method. For a fixed assignment of particle and hole lines, we first impose all inequality constraints generated by the particle–hole structure of the diagram, which determines which vertices must occur later in time than others. We then construct the time ordering sequentially: at each step, we choose uniformly among the vertices that are not blocked by any active inequality (i.e., those that are not required to come after another unplaced vertex). This procedure guaranties that the generated ordering automatically satisfies all particle–hole constraints and produces only non-vanishing Goldstone contributions.

\subsection*{Normalization sector}

\begin{figure}[H]
    \centering
    \includegraphics[width=0.24\linewidth]{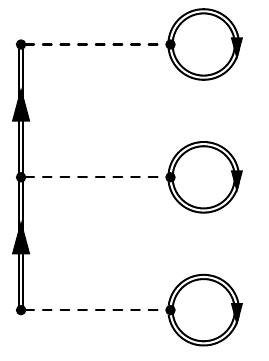}
    \caption{Normalization diagram at order three.}
    \label{fig:three_loops}
\end{figure}

Eq. \eqref{eqDiagMC} depends on the normalization factor $Z_{n_1n_2}^{(tlj)}$, which is unknown. It can, however, be estimated by considering a smaller subset of diagrams, which we call ``normalization diagrams" \cite{Brolli2025, VanHoucke2019}. This subset needs to be simple enough so that its weight $\tilde{Z}_{n_1n_2}^{(tlj)}$ can be computed numerically. For a given order $\nu$, we define the normalization diagram as the diagram with $\nu$ tadpoles, which has the advantage of having a weight function, as defined in \eqref{eqWeight} that scales very favorably as $\sim e_{max}^{7} + \nu e_{max}^6 \sim N_{sp}^{7/3} + \nu N_{sp}^2$. Fig. \ref{fig:three_loops} shows the normalization diagram of order three.

If, during the simulation, the normalization diagram is visited $\mathcal{N}$ times, in the limit $N \to \infty$ we have  $\mathcal{N} / N = \tilde{Z}_{n_1n_2}^{(tlj)} / Z_{n_1n_2}^{(tlj)}$.
We can calculate the normalization factor as 

\begin{equation}
    Z_{n_1n_2}^{(tlj)} = \frac{N}{\mathcal{N}} \tilde{Z}_{n_1n_2}^{(tlj)}.
\end{equation}

The estimation of the normalization factor can be performed very efficiently, as counting the number of visits of the normalization sector is sign-problem-free by definition.

\subsection*{Updates}

The diagrams are sampled with a Markov chain, and to reproduce the probability distribution function \eqref{eqWeight} we use the Metropolis-Hastings algorithm \cite{Hastings1970, Metropolis1953}. The role of the proposal function at each walker step is played by local updates that propose to modify the quantum numbers and topology of the diagrams. We have repeated different simulations to sample each order, and at each perturbative order, we have used the following updates

\begin{figure}[H] 
    \centering
    \includegraphics[width=1\linewidth]{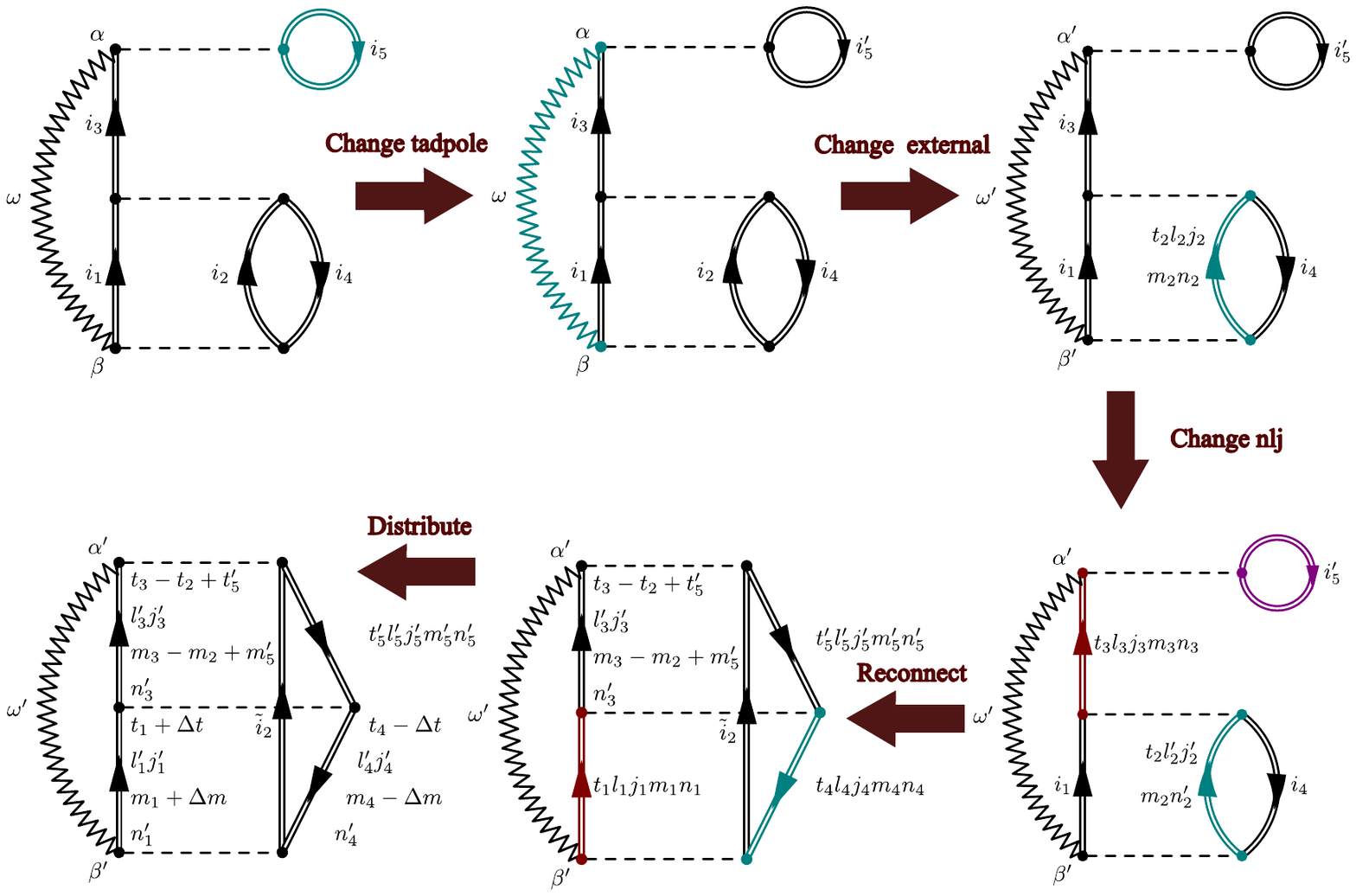}
    \caption{Chain of all the updates acting sequentially on a diagram. The zigzag line connecting the incoming and outgoing vertices is referred to as the ``measuring line."}
    \label{fig::updatesSM}
\end{figure}

To explain the updates, we always refer to Fig. \ref{fig::updatesSM}.

\begin{itemize}
    
    \item \textbf{Change tadpole}

    The update is proposed only if the current diagram contains at least one tadpole. It consists of changing the quantum number flowing through one of the tadpole propagators. The tadpole to be modified is selected uniformly among all existing tadpoles, and the new quantum number is drawn uniformly from the full set of orbitals.

    The acceptance ratio for this move is

    \begin{equation}
        q_{CT} = \mathrm{min} \left( 1,  \left\lvert \frac{V_{\alpha i_5 i_3 i_5}}{V_{\alpha i_5' i_3 i_5'}} \right\rvert \right).
    \end{equation}

    \noindent Here, $V$ denotes the interaction vertex attached to the tadpole. The index $i_5$ corresponds to the original quantum number of the tadpole propagator, while $i_5'$ denotes the newly proposed one. We refer to figure \ref{fig::updatesSM}.

    \item \textbf{Change external}

    The update proposes to change the principal quantum number of the external quantum numbers. The partial wave is kept fixed because we perform separate simulations for each one.
    The acceptance ratio of the update is

    \begin{equation}
        q_{CE} = \mathrm{min} \left( 1, \frac{W(\omega')}{W(\omega)}  \left\lvert \frac{V_{\alpha' i_5' i_3 i_5'} V_{i_1 i_2 \beta' i_4}}{V_{\alpha i_5' i_3 i_5'} V_{i_1 i_2 \beta i_4}} \right\rvert \right),
    \end{equation}

    \noindent where $W(\omega)$ is the Gauss-Legendre weight of the node.

    \item \textbf{Change nlj}

    The update selects a non-tadpole propagator with a uniform probability distribution. If the propagator has a magnetic quantum number $m$, the update proposes to select a new $j$, which we call $j'$, between $\lvert m \rvert$ and $e_{max} + \frac{1}{2}$. The new value of $l$, which we refer to as $l'$, is chosen to be $\frac{1}{2} \lvert 2j \pm 1\rvert$, where $\pm$ is selected so that the new value of $l$ has the same parity as the old value (or the diagram would be zero due to parity conservation). 
    If the update selects $j' = 2e_{max} + 1$, and $l$ is even, the proposed diagram has $l' = e_{max} + 1$, which is outside the model space. In this scenario, the update is simply rejected. Defining $n(l) = \left\lfloor \frac{e_{max} - l'}{2} \right\rfloor$, the principal quantum number is chosen uniformly in $\left[0, n(l')  \right]$.

    The acceptance ratio of the update is 

    \begin{equation}
        q_{Cnlj} = \mathrm{min} \left( 1, \frac{n(l')}{n(l)} \left\lvert \frac{V_{i_3i_4i_1i_2'} V_{i_1i_2'\beta'i_4}}{V_{i_3i_4i_1i_2} V_{i_1i_2\beta'i_4}} \right\rvert \right).
    \end{equation}

    \item \textbf{Reconnect}

    The update proceeds by selecting uniformly a non-tadpole propagator, which necessarily connects two distinct vertices. It then proposes to interchange the endpoints of two other propagators attached to these two vertices, leaving the selected propagator unchanged. If any of the propagators affected by this interchange corresponds to the measuring line, the proposal is immediately rejected.
    
    After the topology change, the quantum numbers of the selected propagator are reassigned so as to preserve all conservation laws, as illustrated in Fig.~\ref{fig::updatesSM}. In particular, the total angular momentum $j$ is kept fixed, i.e. $j_3' = j_3$. The orbital angular momentum $l_3'$ is chosen such that it is compatible with the fixed $j_3$ and ensures parity conservation at both vertices. The radial quantum number $n_3'$ is then drawn uniformly in the interval $[0, n(l_3')]$. The magnetic and isospin projections are determined according to the constraints shown in Fig.~\ref{fig::updatesSM}. If any of these assignments violates the allowed ranges (i.e. produces an overflow or underflow), such as $t_3' > 1$, the proposal is rejected. 

    The acceptance ratio is 

    \begin{equation}
        q_{R} = \mathrm{min} \left( 1, \frac{N_T'}{N_T} \frac{\mathcal{R}(\mathcal{T'})}{\mathcal{R}(\mathcal{T})} \left\lvert \frac{ V_{\alpha' i_5' i_3' \tilde{i}_2} V_{i_3'i_4i_1i_5'}}{ V_{\alpha' i_5' i_3 i_5'} V_{i_3i_4i_1\tilde{i}_2}} \right\rvert \right),
    \end{equation}

    \noindent where $N_T'$ and $N_T$ are the number of tadpoles in the new, and old diagrams respectively.

    \item \textbf{Distribute}

    This update is called only if the diagram contains no tadpoles. It selects two propagators connecting the same pair of vertices and proposes to redistribute their quantum numbers.
    
    Referring to Fig.~\ref{fig::updatesSM}, we first draw shifts $\Delta t \in [-1,1]$ and $\Delta m \in [-1,1]$. We then propose new orbital angular momenta $l_1'$ and $l_4'$, each differing by at most $\pm 1$ from their original values, while remaining within the model space. The proposed values are additionally constrained to preserve parity at the vertices. The corresponding total angular momenta $j_1'$ and $j_4'$ are then selected at random from the allowed set $\left\{ |l - \tfrac{1}{2}|, l + \tfrac{1}{2} \right\}$.
    
    If any of the proposed quantum numbers lies outside its admissible range (for example, if $m < -j$ or $m > j$), the update is rejected. The radial quantum numbers $n_1'$ and $n_4'$ are drawn uniformly from the intervals $[0, n(l_1')]$ and $[0, n(l_4')]$, respectively.
    
    Note that when $l' = 0$, the allowed values of $j'$ reduce to the single possibility $j' = \tfrac{1}{2}$, whereas for $l' \neq 0$ there are two possible values. This asymmetry must be accounted for in the acceptance probability. To this end, we define

    \begin{equation}
        C_j(l) = \begin{cases}
        1 \; \mathrm{if} \; l = 0\\
        2 \; \mathrm{if} \; l \ne 0
        \end{cases}
    \end{equation}

    We also introduce a combinatorial factor $C_l(l \to l')$, defined as the number of distinct ways to propose the transition from $l$ to $l'$.

    The acceptance ratio of the update is given by

    \begin{equation}
        q_{D} = \mathrm{min} \left( 1, \frac{C_l(l_1 \to l_1') C_l(l_4 \to l_4') C_j(l'_1)C_j(l_4')n(l_1') n(l_4')}{C_l(l_1' \to l_1) C_l(l_4' \to l_4)C_j(l_1)C_j(l_4)n(l_1) n(l_4)} \left\lvert \frac{V_{i_3'i_4'i_1'i_5'}V_{i_1'\tilde{i}_2\beta'i_4'}}{V_{i_3'i_4i_1i_5'}V_{i_1\tilde{i}_2\beta'i_4}} \right\rvert \right).
    \end{equation}
    
\end{itemize}

\subsection*{Galitskii-Migdal-Koltun sum rule}

The Galitskii--Migdal--Koltun (GMK) sum rule \cite{Migdal1958, Koltun1974} allows computing the ground-state energy from the single-particle Green's function,
\begin{equation} \label{eq:GMK}
    E_0^A = \sum_{\alpha \beta} \int_{-\infty}^{E_F} \frac{d\omega}{2\pi}
    \left( H^{(0)}_{\alpha \beta} + \omega \delta_{\alpha \beta} \right)
    \mathrm{Im}\, G_{\alpha \beta}(\omega).
\end{equation}

For a finite regulator $\eta$, this expression becomes divergent as the cutoff at the Fermi energy prevents the cancelation of the Lorentzian tails in integrals of the form
\begin{equation}
    \int_{-\infty}^{E_F} d\omega \, \omega \, \mathrm{Im}\, G_{\alpha\alpha}(\omega),
\end{equation}
and a renormalized formulation is required.

We start from the decomposition of the Green's function into hole and particle spectral functions \cite{Dickhoff:manybody},
\begin{equation} \label{eq:S_def}
\begin{aligned}
    S^h_{\alpha \beta}(\omega) &= \frac{1}{\pi} \mathrm{Im}\, G_{\alpha \beta}(\omega)\, \theta(E_F - \omega), \\
    S^p_{\alpha \beta}(\omega) &= -\frac{1}{\pi} \mathrm{Im}\, G_{\alpha \beta}(\omega)\, \theta(\omega - E_F).
\end{aligned}
\end{equation}
Using Eq.~\eqref{eq:S_def}, Eq.~\eqref{eq:GMK} can be rewritten as
\begin{equation}
    E_0^A = \frac{1}{2} \sum_{\alpha \beta} \int_{-\infty}^{+\infty} d\omega \,
    \left( H^{(0)}_{\alpha \beta} + \omega \delta_{\alpha \beta} \right)
    S^h_{\alpha \beta}(\omega).
\end{equation}

From the Dyson equation, the asymptotic behavior of the Green's function is
\begin{equation}
    \mathrm{Im}\, G(\omega) \xrightarrow[\omega \to \infty]{} 
    \mathrm{Im}\, G^{(\infty)}(\omega) + \mathcal{O}\!\left(\frac{1}{\omega^4}\right),
\end{equation}
where $G^{(\infty)}$ is the propagator constructed from the static part of the self-energy. Since $G^{(\infty)}$ is known analytically from its Lehmann representation, its contribution can be evaluated analytically.

We split the spectral function as
\begin{equation}
    \delta S^h_{\alpha \beta}(\omega) = S^h_{\alpha \beta}(\omega) - S^{(\infty),h}_{\alpha \beta}(\omega),
\end{equation}
and rewrite the energy as
\begin{equation} \label{eq:GMK_split}
\begin{aligned}
    E_0^A &= \frac{1}{2} \sum_{\alpha \beta} \int_{-\infty}^{+\infty} d\omega \,
    \left( H^{(0)}_{\alpha \beta} + \omega \delta_{\alpha \beta} \right)
    \delta S^h_{\alpha \beta}(\omega) \\
    &\quad + \frac{1}{2} \sum_{\alpha \beta} \int_{-\infty}^{+\infty} d\omega \,
    \left( H^{(0)}_{\alpha \beta} + \omega \delta_{\alpha \beta} \right)
    S^{(\infty),h}_{\alpha \beta}(\omega).
\end{aligned}
\end{equation}
The second term is computed analytically, while the first one is convergent since the integrand decays at least as $1/\omega^3$. This remains true for finite $\eta$.

In practice, $S^h_{\alpha \beta}(\omega)$ is not directly accessible for finite $\eta$, due to the smearing across the Fermi surface. To avoid this issue, we use the sum rules \cite{Dickhoff:manybody}
\begin{equation}
\begin{aligned}
    \int_{-\infty}^{+\infty} d\omega \, \left(S^h_{\alpha \beta} + S^p_{\alpha \beta}\right)
    &= \delta_{\alpha \beta}, \\
    \int_{-\infty}^{+\infty} d\omega \, \omega \left(S^h_{\alpha \beta} + S^p_{\alpha \beta}\right)
    &= h_{\alpha \beta} + \Sigma^{(\infty)}_{\alpha \beta}.
\end{aligned}
\end{equation}
Using eq. \eqref{eq:S_def}, this implies
\begin{equation}
\begin{aligned}
    \int_{-\infty}^{+\infty} d\omega \, \delta S^h_{\alpha \beta}(\omega)
    &= \frac{1}{2\pi} \int_{-\infty}^{+\infty} d\omega \, \mathrm{Im}\, \delta G_{\alpha \beta}(\omega), \\
    \int_{-\infty}^{+\infty} d\omega \, \omega \, \delta S^h_{\alpha \beta}(\omega)
    &= \frac{1}{2\pi} \int_{-\infty}^{+\infty} d\omega \, \omega \, \mathrm{Im}\, \delta G_{\alpha \beta}(\omega),
\end{aligned}
\end{equation}
with $\delta G = G - G^{(\infty)}$.

The GMK sum rule can thus be recast as
\begin{equation} \label{eq:GMK_final}
\begin{aligned}
    E_0^A &= \frac{1}{4\pi} \sum_{\alpha \beta} \int_{-\infty}^{+\infty} d\omega \,
    \left( H^{(0)}_{\alpha \beta} + \omega \delta_{\alpha \beta} \right)
    \mathrm{Im}\, \delta G_{\alpha \beta}(\omega) \\
    &\quad + \frac{1}{2} \sum_{\alpha \beta} \int_{-\infty}^{+\infty} d\omega \,
    \left( H^{(0)}_{\alpha \beta} + \omega \delta_{\alpha \beta} \right)
    S^{(\infty),h}_{\alpha \beta}(\omega),
\end{aligned}
\end{equation}
which is finite for $\eta \neq 0$.

\bibliography{bibliography}

\bibliographystyle{apsrev4-2}